# Magnetic Moment Softening and Domain Wall Resistance in Ni Nanowires

J. D. Burton[1,3], R. F. Sabirianov[2,3], S. S. Jaswal[1,3], and E. Y. Tsymbal[1,3*]

[1]*Department of Physics and Astronomy, University of Nebraska, Lincoln, Nebraska 68588-0111, USA*
[2]*Department of Physics, University of Nebraska, Omaha, Nebraska 68182-0266, USA*
[3]*Center for Materials Research and Analysis, University of Nebraska, Lincoln, Nebraska 68588-0111, USA*

O. N. Mryasov

*Seagate Research, Pittsburgh, Pennsylvania 15222, USA*

Magnetic moments in atomic scale domain walls formed in nanoconstrictions and nanowires are softened which affects dramatically the domain wall resistance. We perform *ab initio* calculations of the electronic structure and conductance of atomic-size Ni nanowires with domain walls only a few atomic lattice constants wide. We show that the hybridization between noncollinear spin states leads to a reduction of the magnetic moments in the domain wall. This magnetic moment softening strongly enhances the domain wall resistance due to scattering produced by the local perturbation of the electronic potential.

DOI:                    PACS number(s):

The study of domain walls (DWs) in bulk ferromagnetic materials began in the early 20th century with contributions from Bloch[1], Landau and Lifshitz[2], and Néel[3]. In bulk 3d metal ferromagnets DWs are wide (~100 nm) on the scale of the lattice spacing due to the strong exchange interaction which tends to align neighboring regions of magnetization, compared to anisotropic effects which prefer the magnetization to lie along specific directions. With the recent interest in magnetic nanostructures it has been shown that DWs can be quite thin due to the enhanced anisotropic effect of constrained geometry in nanowires and nanoconstrictions.[4] For example, Prokop *et al.*[5] have recently observed DWs of only a few lattice constants wide in single monolayer Fe nanowires grown on Mo.

The DW width controls the DW resistance. The origin of the DW resistance is known to be the mixing of up- and down-spin electrons due to the mistracking of the electron's spin in passing through the DW.[6] The narrower DW width results in a larger angle between the magnetization directions of successive atomic layers thereby lowering the electron transmission. In bulk ferromagnets DWs do not affect appreciably the resistance because the DW wall width is much larger than the Fermi wave length, and hence electrons can follow adiabatically the slowly varying magnetization direction within the DW. In magnetic nanoconstrictions, where an atomic scale DW can be formed between two electrodes magnetized antiparallel to one another, the DW resistance may be appreciable. The ballistic transport across such a DW leads to interesting magnetoresistive phenomena.[7,8]

The theoretical description of the DW resistance in the ballistic transport regime has attracted much attention. The approaches which were used are based on either free-electron models[9,10,11,12] in which the DW is represented by an appropriate potential profile or first-principles calculations[13,14,15,16] in which the DW is typically described by a spin-spiral structure. All these models assume that the DW is *rigid*, i.e. they neglect any spatial variation of the magnitude of the magnetic moment across the DW. It is well established, however, that the magnitude of the magnetic moments in itinerant magnets can depend strongly on the orientation of the neighboring moments.[17] This effect is relatively weak in well localized ferromagnets like Fe, but can reduce and even destroy the atomic magnetic moment of the itinerant ferromagnets, like Ni.

The origin of this phenomenon is the presence of hybridization between noncollinear spin states. In the uniformly magnetized material with no spin-orbit coupling the minority and majority spin bands are independent. However, in a noncollinear state, such as a DW, this is no longer the case and the two spin bands are hybridized. This spin mixing leads to charge transfer and level broadening, which results in the reduction of the overall exchange splitting between majority and minority states on each atom, and hence the atomic moments are reduced. In bulk DWs of any ferromagnetic material, this effect is small due to the slow variation of the direction of the magnetization. For the atomic scale DWs formed in magnetic nanostructures, however, there is a large degree of canting between neighboring magnetic moments. This results in a significant hybridization between spin states which leads to a reduction, or *softening*, of the magnetic moments within the constrained DW. The magnetic moment softening affects the DW resistance due to the local perturbation in the electronic potential.

We note that the spatial variation of the magnetization in DWs has been suggested before, but only in the context of finite temperature magnetic disorder of the (fixed magnitude) atomic magnetic moments in a DW.[18,19,20] The effect of spatial variation of the magnetization on DW resistance was addressed previously, but only within the diffusive transport regime and a free electron model.[21]

In this Letter we illustrate the importance of magnetic moment softening by performing *ab initio* calculations of the electronic structure and ballistic conductance of atomic scale DWs in Ni nanowires. We show that the magnetic moments within DWs only a few lattice constants wide can be significantly reduced compared to the magnetic moments in a uniformly magnetized wire due to the presence of substantial hybridization between spin states. We find that the magnetic moment softening strongly enhances the DW resistance due to additional scattering resulting from the local perturbation in the electronic potential.

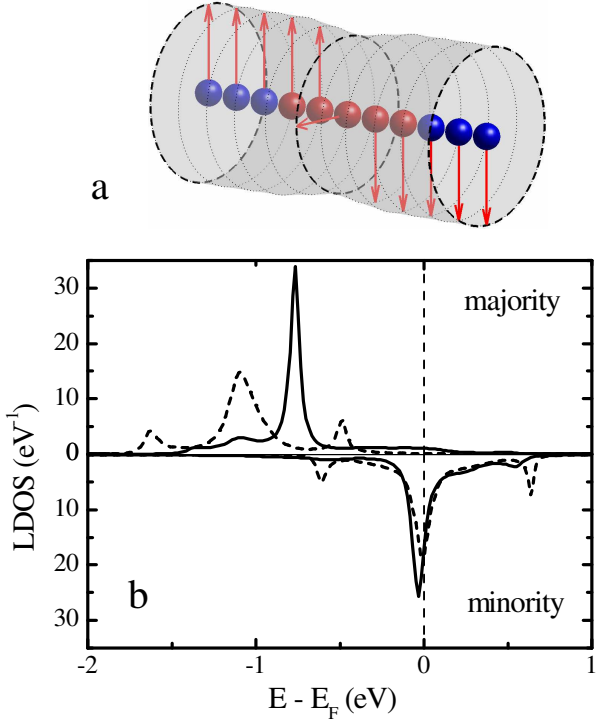

**Fig. 1** (Color online) (a) A monatomic wire with an $N = 1$ DW. The arrows show the orientation and relative magnitude of calculated magnetic moments. (b) DOS per atom of the uniformly magnetized monatomic wire (dashed curve) and the LDOS at the center of an $N = 1$ DW (solid curve).

Density functional calculations of the spin-dependent electronic structure of atomic scale DWs in Ni nanowires were performed using the tight-binding linear muffin-tin-orbital method[22] in the atomic sphere approximation and the local spin density approximation for the exchange-correlation energy. We used the real space recursion method[23] with a Beer-Pettifor terminator[24] to calculate the local density of states (DOS). Ultrathin domain walls were examined for two different free-standing Ni wires based on the bulk fcc structure: (110) monatomic chains and 5×4 wires (described below). The wires were surrounded by a few layers of empty spheres to accurately describe the charge density. All the structures used the lattice constant $a$ = 3.52 Å of bulk fcc Ni.

In the calculations we constructed a central region containing the DW, surrounded by two uniformly magnetized leads aligned antiparallel relative to each other. The self-consistent calculations were carried out for the central region and a large enough portion of the surrounding leads so that the outermost atoms of the self-consistent region were similar to those in the remaining semi-infinite section of the leads in the uniformly magnetized state. A DW was modeled by a finite spin spiral with the relative angle between neighboring atomic layers of magnetic moments being $180°/(N + 1)$, where $N$ is the number of atomic layers in the DW. In all the calculations the orientation of the magnetic moments was held fixed.

First we consider an $N = 1$ DW in a monatomic Ni wire (Fig 1a). Fig. 1 displays the results of our calculations for the $N = 1$ DW. The electronic potentials of the antiparallel magnetized lead sites (blue) were frozen, while on the central five sites (red) the electronic structure was calculated self-consistently. We find a 16% reduction of the magnetic moment on the central site and a 7% reduction on the two neighboring sites. The reduction in the magnetic moment on the central site is due to the hybridization with the different spin states of the non-collinear neighbors which results in the reduction of the exchange splitting, as described earlier. This fact is evident from Fig. 1b which shows the local DOS for the uniformly magnetized monatomic Ni wire and for the central atom of an $N = 1$ DW. The reduction in the moment of the nearest neighbors of the central atom is only about half of that of the central atom because they are noncollinear with only one of the two neighbors. Also apparent in Fig. 1b is an overall broadening of the DOS on the central site compared to the uniformly magnetized state produced by the spin mixing.

Fig. 2 shows the magnetic moment profile for several DW widths in the monatomic wire. The reduction in the moment is strongly dependent on the width of the DW. For $N = 0$ and $N = 1$ DWs the effect of softening is the largest owing to the fact that the degree of noncollinearity is the largest in these two cases. As the width of the DW increases the softening decreases because of the reduction of the angle between the nearest neighbor magnetic moments.

Fig. 3a shows the 5×4 wire which has five atoms in one layer and four in the next layer resulting in three

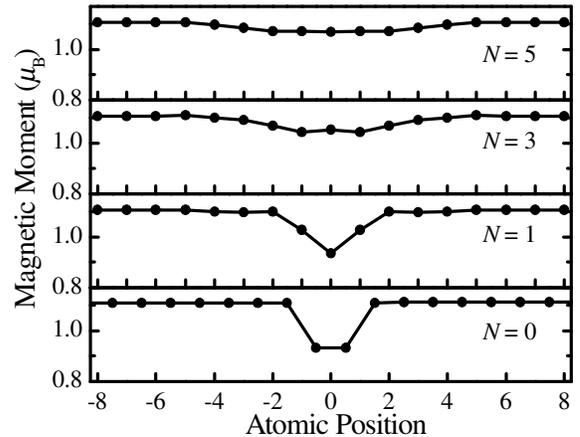

**Fig. 2** Magnetic moments in the monatomic Ni wire as a function of distance from the DW center for several DW widths. The $N = 1$ plot corresponds to Fig. 1a.

determine the contribution of magnetic moment softening to the DW resistance.

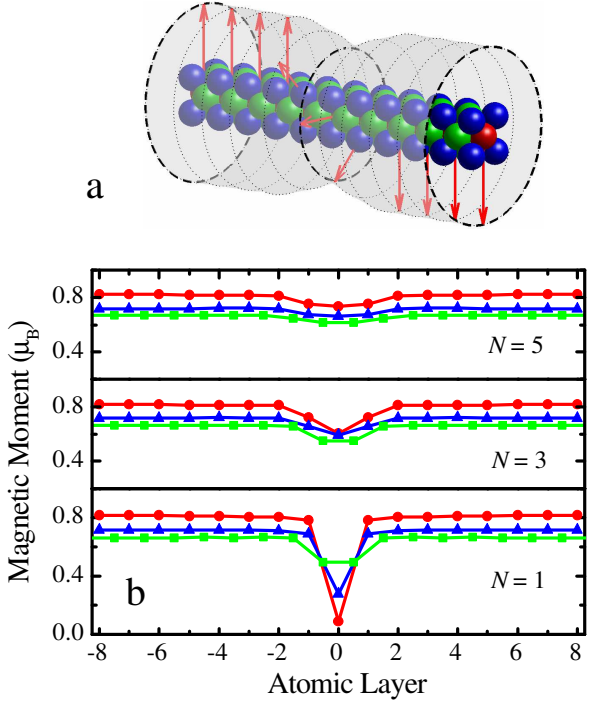

**Fig. 3** (Color Online)   (a) The 5×4 nanowire showing the three nonequivalent sites and an $N = 3$ DW.  Each arrow represents the magnitude and orientation of the average magnetic moment of the plane with 5 atoms. (b) The self consistent magnetic moments for several DW widths.  The color in the plot corresponds to the site of the same color in (a).

nonequivalent sites, each indicated by a different color. Although the average magnetic moment of the uniformly magnetized 5×4 wire,  $\mu = 0.7\mu_B$, is lower than that of the monatomic wire, $\mu = 1.1\mu_B$, we find that in the presence of ultra-thin DWs the spatial variation of the magnetization displays qualitatively similar behavior.  Fig. 3b shows the magnetic moments found for three DW widths in the 5×4 wire.  The largest reduction in magnetic moment is nearly 90% for the central site of the $N = 1$ DW, and the effect is still substantial for the $N = 5$ DW where the magnetic moment of the central layer is softened by about 10%.  The effect is significantly larger in the 5×4 wire than that in the monatomic chain due to the enhanced hybridization reflecting an increase in the number of neighboring atoms.

The magnetic moment softening affects dramatically the conductance across DWs.  We calculate the conductance of the 5×4 Ni wire with an $N = 1$ DW using the standard tight-binding technique described in detail in Ref. 16.  It should be noted, however, that in Ref. 16 the Hamiltonian for the DW region was built by simply rotating (in spin space) the self-consistent potentials obtained for the uniformly magnetized wire.  Here we derive the Hamiltonian from the self-consistent potentials found in the presence of the $N = 1$ DW, i.e. taking into account magnetic moment softening.  For comparison we also calculate the conductance in the same way as Ref. 16 which allows us to

Fig. 4a shows the spin-resolved conductance, $G_{FM}$, of a uniformly magnetized 5×4 Ni wire as a function of energy $E$.  The conductance is quantized in steps of $e^2/h$, corresponding to the number of bands crossing the energy $E$.  The bands are seen in Fig. 4b as being bound by the sharp peaks corresponding to the band edges.  The conductance in the zero-bias limit is given by the value at the Fermi energy, $E_F$, located at the center of the narrow minority band.  For the 5×4 Ni wire it is 14 $e^2/h$.  Fig. 4c displays the conductance through an $N = 1$ DW, $G_{DW}$.  For a "rigid" DW (no softening of the magnetic moments) the Hamiltonian is constructed from the self-consistent potentials of the uniformly magnetized wire, while for the "soft" DW we use the potentials that yield reduced magnetic moments in the DW region, as seen in Fig. 3.  We find that the magnetic moment softening in the DW leads to a reduction in the conductance at the Fermi energy from 5.38 $e^2/h$ to 3.21 $e^2/h$.  This implies the significant enhancement of the DW magnetoresistance, $(G_{FM} - G_{DW})/G_{DW}$, from 160% to about 340%.

The origin of this phenomenon can be understood using a simple tight-binding model of a monatomic chain with an abrupt $N = 0$ DW.  We model the majority states by a wide band, characterized by the (large) nearest neighbor hopping parameter $t_{maj}$, to emulate the wide band near the Fermi energy (see Fig. 4b).  The minority states are modeled by a narrow band with (small) hopping $t_{min}$, offset from the

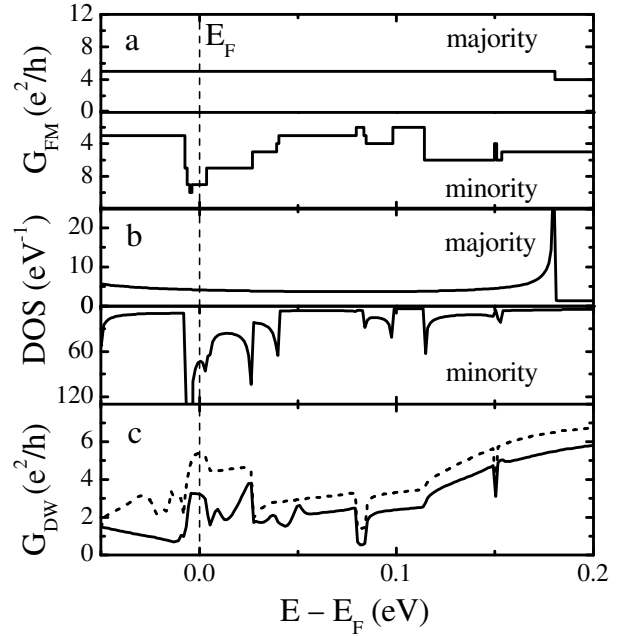

**Fig. 4** (a) Spin-resolved conductance, $G_{FM}$, and (b) density of states near the Fermi energy, $E_F$, as a function of energy for the uniformly magnetized 5×4 Ni wire. (c) Conductance of the $N = 1$ DW, $G_{DW}$, as a function of energy for the rigid DW (dashed curve) and for the soft DW (solid curve).

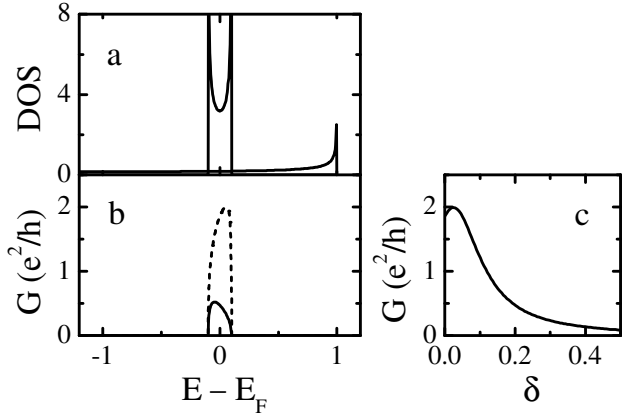

**Fig. 5** Results of a simple tight-binding model. (a) DOS of the uniformly magnetized state versus energy for majority- (wide band) and minority- (narrow band) spin electrons. (b) Conductance versus energy for a rigid DW (δ = 0, dashed line) and for a soft DW (δ = 0.2, solid line). (c) Conductance versus δ for $E = E_F$.

center of the majority band by energy Δ. To model magnetic moment softening we assume that the onsite energies of the two interface sites are shifted relative to the uniformly magnetized leads. Majority states are shifted up by energy δ and minority states are shifted down by the same amount δ. This corresponds effectively to a reduction of the local exchange splitting by 2δ on the two interface sites, and hence softening of the magnetic moment. The parameters for our model are chosen as follows: $t_{maj} = 1$, $t_{min} = 0.05$, $\Delta = 1$. The results of the model are shown in Fig. 5. As is seen from Fig. 5b,c, the model predicts a drastic reduction in the conductance as a result of magnetic moment softening. The origin of this reduction is the local perturbation in the electronic potential which leads to stronger scattering of transport electrons by the DW. When the onsite energies of the interface atoms are shifted with respect to the narrow minority band, which determines the energy window for conductance, these sites act as additional scatterers that hinder conductance.

In conclusion, we have shown that in atomic scale domain walls of Ni nanowires the magnetic moments are softened due to the noncollinearity with the neighboring magnetic moments. This effect is stronger in wires of larger cross-section due to the enhanced hybridization but falls off quickly with increasing DW width due to a decreasing degree of noncollinearity. The magnetic moment softening significantly enhances the DW resistance as a result of scattering produced by the local perturbation of the electronic potential.

This work is supported by Seagate Research, the NSF (Grant Nos. DMR-0203359 and MRSEC: DMR-0213808), and the Nebraska Research Initiative. The calculations were performed using the Research Computing Facility of the University of Nebraska-Lincoln.

* Electronic address: tsymbal@unl.edu